%% file: noc.tex
\documentclass[tighten, onecolumn, trackchanges, 12pt]{aastex62}
\usepackage{listings}
\usepackage{color}
\usepackage{graphicx,times}
\usepackage{natbib}
\usepackage{amssymb}
\usepackage{newtxtext}
\usepackage[T1]{fontenc}
\usepackage{ae,aecompl}
\usepackage{threeparttable}
\usepackage{ulem}
\usepackage{txfonts}
\usepackage{amssymb}	

\def\kms  {km~s$^{-1}$}

\bibpunct{(}{)}{;}{a}{}{,}
\usepackage{hyperref}
\usepackage{longtable}
\usepackage{booktabs}
\definecolor{dkgreen}{rgb}{0,0.6,0}
\definecolor{gray}{rgb}{0.5,0.5,0.5}
\definecolor{mauve}{rgb}{0.58,0,0.82}
\definecolor{golden}{rgb}{0.86,0.65,0.01}

\usepackage{CJK}
\usepackage{graphicx}
\usepackage{hyperref}  
\begin{document}
\begin{CJK*}{UTF8}{gbsn}
\received{11 Jul~2022}
\revised{30 Sep~2022}
\accepted{30 Sep~2022}
\title[]{Unveiling hidden stellar aggregates in the Milky Way: 1656 new star clusters found in Gaia EDR3}
\correspondingauthor{Zhihong He}
\email{hezh@cwnu.edu.cn}
\author[0000-0002-6989-8192]{Zhihong He (何治宏)}
\affil{School of Physics and Astronomy, China West Normal University, No. 1 Shida Road, Nanchong 637002, China }
\author{Xiaochen Liu (刘效臣)}
\affil{School of Physics and Astronomy, China West Normal University, No. 1 Shida Road, Nanchong 637002, China }
\author{Yangping Luo (罗杨平)}
\affil{School of Physics and Astronomy, China West Normal University, No. 1 Shida Road, Nanchong 637002, China }
\author{Kun Wang (王坤)}
\affil{School of Physics and Astronomy, China West Normal University, No. 1 Shida Road, Nanchong 637002, China }
\author{Qingquan Jiang (蒋青权)}
\affil{School of Physics and Astronomy, China West Normal University, No. 1 Shida Road, Nanchong 637002, China }
\vspace{10pt}
\begin{abstract}
We report 1,656 new star clusters found in the Galactic disk (|b|<20~degrees) beyond 1.2~kpc, using Gaia EDR3 data. Based on an unsupervised machine learning algorithm, DBSCAN, and followed our previous studies, we utilized a unique method to do the data preparation and obtained the clustering coefficients, which proved to be an effective way to search blindly for star clusters. We tabulated the physical parameters and member stars of the new clusters, and presented some interesting examples, including a globular cluster candidate. The cluster parameters and member stars are available at CDS via anonymous ftp to \url{https://cdsarc.cds.unistra.fr/ftp/vizier.submit//he22c}. We examined the new discoveries and discussed their statistical properties. The proper motion dispersions and radii of star clusters are the same as the previously reported. The new clusters beyond 1.2~kpc are older than those in the solar neighborhood, and the new objects found in the third Galactic quadrant present the lowest line-of-sight extinctions. Combined with our previous results, the total population of new clusters and candidates detected through our method is 2,541, corresponding to 55\% of all newly published clusters in the Gaia era. The number of cataloged Gaia star clusters was also increased to nearly six thousand. In the near future, it is necessary to make an unified confirmation and member star determination for all reported clusters.
\end{abstract}

\section{Introduction}\label{sec:intro}
In recent years, searching for star clusters (hereafter SCs) in Gaia's~\citep[][]{Gaia16} stellar astrometric data has been an attractive subject for researchers. After the second data release~\citep[][DR2]{Gaia18-Brown}, about 1,200 pre-Gaia open clusters (OCs) have been reidentified~\citep[][]{CG18}, and more than 1,500 new ones have been unveiled for the first time~\citep[e.g.][]{Liu19,Sim19,CG19-0,Castro19,Kounkel20,Castro20,Ferreira20,ferreira21,he21,he22a,Hunt21}. Based on the early third data release~\citep[][EDR3]{gaia2021}, researchers have reported more than ~1,500 new OCs and candidates~\citep[][]{castro22,hao22,he22b} so far, which covered the distance range from 100~pc to 5~kpc. 

New SCs are constantly being discovered, mainly for two reasons. On the one hand, it is due to the continuous improvement of the accuracy of Gaia's astrometric parameters. For example, for five-parameter astrometry, the typical uncertainties of the parallax and proper motion for stars about 17~mag for DR2 are 0.1~mas and 0.2~mas/yr~\citep[][]{Gaia18-Brown}, and for EDR3 are 0.07~mas and 0.07~mas/yr~\citep[][]{gaia2021}, respectively. 
This improvement in astrometric accuracy helps to distinguish cluster members from stars along the same line of sight, which allows one to remove background and foreground stars by performing parallax cuts, and field stars with similar parallaxes can be identified through their dissimilar proper motion magnitudes and directions. 
On the other hand, clustering methods used to search for SCs are diverse and are constantly being refined. In particular, DBSCAN, an algorithm with a history of more than two decades~\citep{Ester96}, has been proved to be the most successful method for blind SC searches~\citep{Cantat22}.

The search efficiency of DBSCAN  is highly dependent on the input clustering parameters~\citep{gao13}: $\xi$ and $MinPts$. For any data point, $P_i$, in vector space, the number of neighbors $n_{point}$ within radius $\xi$  of $P_i$ will be calculated through this algorithm, and $P_i$ will be marked as one of three possible results: core ($n_{point} \geqslant MinPts$), member ($n_{point} < MinPts$, but the point is a neighbor of a core point), or noise ($n_{point} < MinPts$, and it is not adjacent to any core). Usually, $\xi$  will be selected according to the histogram's characteristics of the $k_{th}$ nearest neighbor distance~\citep[e.g.][$kNND$]{Ester96,Castro18,xusk19}, and $MinPts$ is an empirical value~\citep{Sander98}, which is usually set between 5 and 25 in cluster studies~\citep[e.g.][]{gao14,Castro20}.

In our previous work~\citep{he21}, we uniquely replaced coordinates and proper motions with linear scales and linear velocities, and we used a new method for computing the $\xi$ parameter. The benefit of these improvements was that field stars and cluster members were more efficiently distinguished. Based on this, we have found about 900 new SC candidates in Gaia DR2 and EDR3, and also found more abundant member stars of the known SCs~\citep[][hereafter H22a, H22b, respectively]{he22a,he22b}. For studies of the evolution of stars and SCs, these new discoveries made the Galactic cluster sample more complete, and for studies of the structure and evolution of the Milky Way, a larger sample allowed one to trace the Galactic structure better over wider spatial and temporal scales.

However, the number of known SCs is still far from the estimated number of expected SCs in the Milky Way. At present, only 3,500 to 4,000 OCs and candidates have been identified from Gaia's astrometric data, which are two orders of magnitude lower than the total estimate~\citep[$\sim10^5$,][]{Piskunov06}. It is reasonable to believe that many undiscovered OCs are still hidden in dense stellar regions.  In this work, using the same methods and steps as H22b, we extended the search to regions beyond 1.2~kpc from the Solar System to find more undiscovered stellar aggregates.

The remainder of this paper is structured as follows. Section 2 describes the data that were used to search for new clusters and the method adopted to detect clusters. Section 3 presents the cluster/member list of the newly discoveries, as well as some examples. Section 4 presents the statistical results and a discussion of the new candidates. Finally, Section 5 summarizes the discoveries and states the prospects of the SC samples.

\section{Data and Method}\label{sec:data}
 \subsection{Stellar data and reference cluster catalogs}\label{sec:cata}	

Our approach focused on searching for SCs that were concentrated in position and motion in the Galactic disk. The subsequent analysis was then based on astrometric data $(l, b, \varpi, \mu_\alpha^\ast, \mu_\delta)$ from Gaia EDR3. We chose |b|<20~degrees sources with an apparent magnitude of G <18~mag as the most suitable database to extract large samples of faint stars in the Galactic disk.  As a continuation of our previous work, we selected a parallax cut of $\varpi$ < 1~mas and an offset of 0.2~mas in order to get all unbiased member stars of clusters near 1.2~kpc, and the groups (clustering result) with average parallax greater than 0.8~mas were removed from the final results. At the same time, the photometric data (G,BP-RP) of Gaia EDR3 were also used to fit isochrones to the newly found cluster candidates.

To build a catalog of new SCs, we cross-matched with previously published sources. The list of references included the works of
~\citet[][]{CG20,Liu19,Ferreira19,Ferreira20,ferreira21,Qin20,he21,he22a,he22b,Hunt21,Casado21,Jaehnig21, dias21, li22,hao22,castro22}. 
In our method we have removed known SCs from the search results, and we also made a roughly cross-match between these published catalogs. For the clusters within the 3$\sigma$ range of $(l, b, \varpi, \mu_\alpha^\ast, \mu_\delta)$ presented in another catalog, we regard them as the matched clusters. According to this estimate, the number of open clusters found in Gaia era was about 3,900. It is necessary to confirm the member stars of all clusters to give more accurate figures. Besides, we also compared with the lists of known clusters in ~\citet{dias21}, newly confirmed pre-Gaia OCs~\citep[][Liu et al. 2022, in prep.]{Dias02,Kharchenko13}, and Galactic globular clusters~\citep{Vasiliev21}.

\subsection{Data preparation}\label{sec:pre}	
Before clustering the stars, we divided the study area into several rectangular-shaped regions that covered the Galactic coordinates with 2 and 6 degrees. For the two-sized subsets, each region has an overlap of 0.5 and 1.5 degrees, respectively. That the two angular sizes used here is because we noticed that the difference between the total number of stars in the region may affect the Gaussian fitting parameters of $kNND$~(Sec.~\ref{sec:clustering}), which made the final number of star cluster members become different. For the same clustering result under these two subsets, we took the group with relatively more members as the final result.

As in our previous work, the input data were $(d\cdot\sin\theta_{l}\cdot\cos b,d\cdot\sin\theta_b,d\cdot\mu_{\alpha^*},d\cdot\mu_{\delta},\varpi)$, and each vector was globally normalized by the standard deviation of the corresponding cluster parameter (see section 2 in H22b). Here, the stellar distance, $d$, was simply taken as the inverse of the parallax. In this approach, we noticed that the distance uncertainties were highly affected by the parallax uncertainties, and hence this approach is accurate only for nearby objects. However, our aim was not to get the precise linear scale and linear velocity of a star; instead, we sought to identify and distinguish stars at different distances in the spatial and proper motion space in the clustering parameter space.

\subsection{Clustering and cross-matching}\label{sec:clustering}

Next, our blind search method in each region was same as that described by H22b, which mainly consisted of three steps. Briefly, in the first step, we computed $\xi$ at different values ​​of $k$ ($k$ = $MinPts$ - 1), where $k$ ranged from [5, 15]. In this step, we fitted the histogram of $kNND$ with the sum of two Gaussian distribution functions (G1 and G2), and used the real solution of G1 = G2 as the value of $\xi$. If there exist distributions with two intersection points, we choose the larger one as the $\xi$. If there was no real solution, we did not perform clustering and instead progressed to the next loop. Then, we performed DBSCAN clustering on the data in each region under different ($k$, $\xi$) values. 
Finally, we combined the results under different ($k$, $\xi$) values, counted the frequency of occurrence of each star, and recorded it as the $N_{re}$ value. 

It should be noticed that, the high-$N_{re}$ stars are located in the center of the $(l, b, \varpi, \mu_\alpha^\ast, \mu_\delta)$ vectors, which should be less contaminated by any possible outliers. However,  we found that in many clusters, the Color-Magnitude positions of some low-$N_{re}$ stars are still located on the isochrone line, so they were still regarded as possible members stars. 
The process were repeated in two-sized subsets, and for the same group, the one with the most members would be taken as the final result.
In the above process, we utilized DBSCAN via \texttt{scikit-learn}~\citep{Pedregosa11} ~\footnote{\url{https://github.com/scikit-learn/scikit-learn}}, and the device we used was a BSCC 64-core cloud computing server.

As mentioned previously, because our main goal was to find new cluster candidates, we cross-matched known clusters from the clustering results. We compared the results in each region with the existing SC catalogs. The method we used was the same as H22b, in which we considered the results that lied within the 3$\sigma$ range of $(l, b, \varpi, \mu_\alpha^\ast, \mu_\delta)$ as matched clusters. In addition, we also performed a visual inspection of each result in the range of 5$\sigma$ in $(l,b)$ to ensure that the candidates found in this work did not duplicate the records in the reference tables. 
Furthermore, we visually inspected the color magnitude diagram (CMD) of each group, and removed those groups, which possess fewer members and do not show the conventional evolutionary sequence expected for SCs.
We finally obtained a set of 2,753 groups, of which 1,097 ones were matched with known star clusters.

\subsection{Isochrone fitting}\label{sec:isochrone}
We performed isochrone fitting for each new result, following the methods used by H22b. It is worth noting that we used the solar metallicity as a fixed isochrone parameter, but in spectroscopy observations, the metallicity of SCs exhibited a gradient as a function of the radius of the Milky Way~\citep{Magrini09}, and at each fixed radius it also changed over a certain range~\citep{Spina21}. However, studies have shown that the change of metallicity had only a small impact on the fitted age and extinction~\citep{Salaris04,cg20arm}. In this work, we chose the range of the distance modulus based on the median parallax of the member stars, and, following the method of~\citet{CG20}, we took random offsets between $(\varpi - \sigma_\varpi)$ to $(\varpi + \sigma_\varpi)$. The steps for the distance modulus and extinction were 0.05 and 0.05~mag, respectively. 

Besides, we noticed that some of the cluster members were obviously unrelated to the expected stellar evolution sequence, which may due to the effects of high reddening and/or field star contamination. Therefore, we performed a visual inspection of each cluster, and for some poor fittings, we used the following method to refit the isochrones: we only picked stars located in the densest parts in the low magnitude end of the CMD. 
At the same time, we manually removed those stars which are obviously far from the evolutionary sequence when doing the fits (mostly from the class 2 clusters in Sec.~\ref{sec:classification}). Since member stars may deviate from the main sequence due to extinction, variable stars or stragglers, we listed all members derived by DBSCAN in the final results.

\section{Results}\label{sec:result}
\begin{figure*}
\begin{center}
	\includegraphics[width=1.\linewidth]{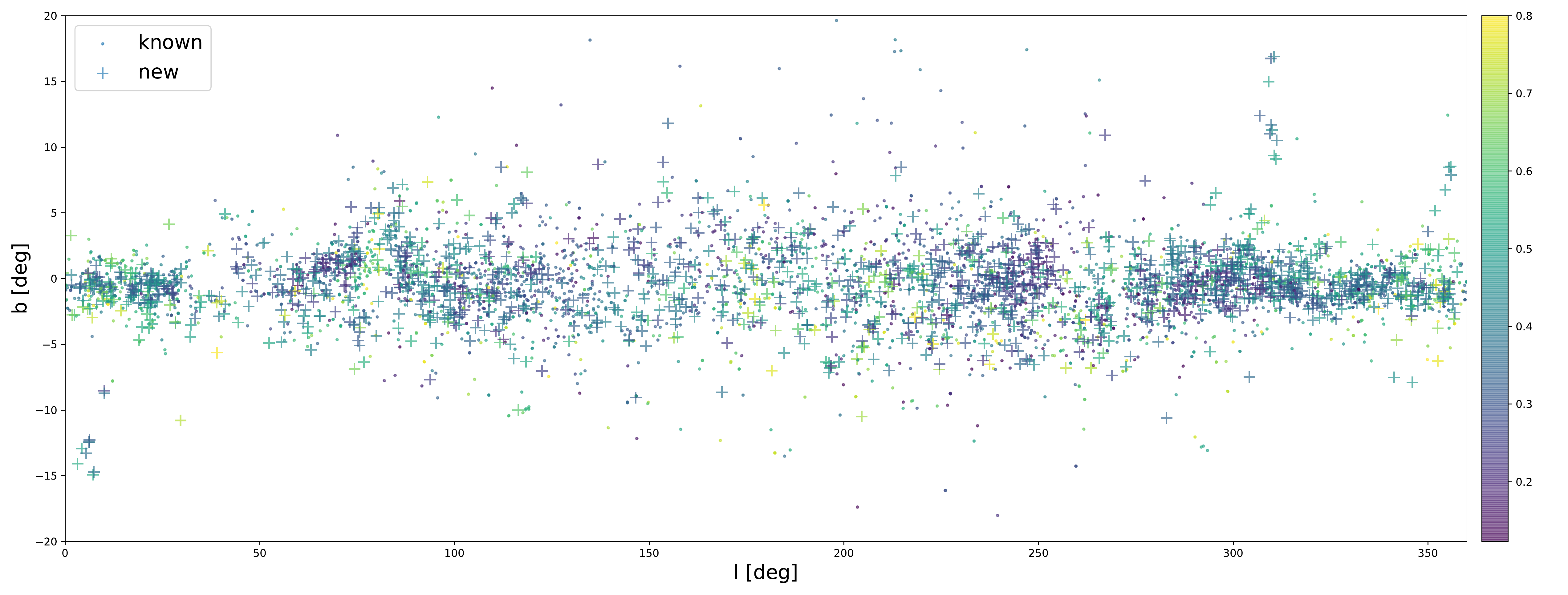}
	\caption{Distribution of the newly found candidates (crosses) and known (dots) OCs ($\varpi$<~0.8~mas) in the Galactic fields. The colors indicate the parallax of each cluster.}
	\label{alloc}
\end{center}
\end{figure*}

After taking the above steps, we obtained a total of 1,656 new SC candidates and a total of 48,714 member stars. The distribution of these candidates in the Galactic coordinate system is shown in Fig.~\ref{alloc}, where more than 90\% of the new candidates are located within 5 degrees of the Galactic plane, and there are fewer new SC candidates found near $l$ to about 40 degrees and 270 degrees, The density in the anti-Galactic center direction was also significantly smaller than that in the Galactic center direction. In this work, the farthest candidate object had a parallax of 0.12~mas and was located in the fourth quadrant. The parameters of the new object catalog are shown in Table~\ref{tab_all}, and their identifications are CWNU~1271 to CWNU~2926. We followed the same approach as H22b, the $(l, b, \varpi, \mu_\alpha^\ast, \mu_\delta)$ values were derived from the median values of the observed data for all member stars, and the corresponding $\sigma$ value represents the dispersion~\footnote{Same as ~\citet{CG20_0}. Throughout this work, the $\sigma$ values are calculated as 1.4826 $\ast$ median absolute deviation, which are robust to outliers.} of each parameter. The full list is available in the electronic version. 

%

\begin{figure*}
\begin{center}
	\includegraphics[width=1.0\linewidth]{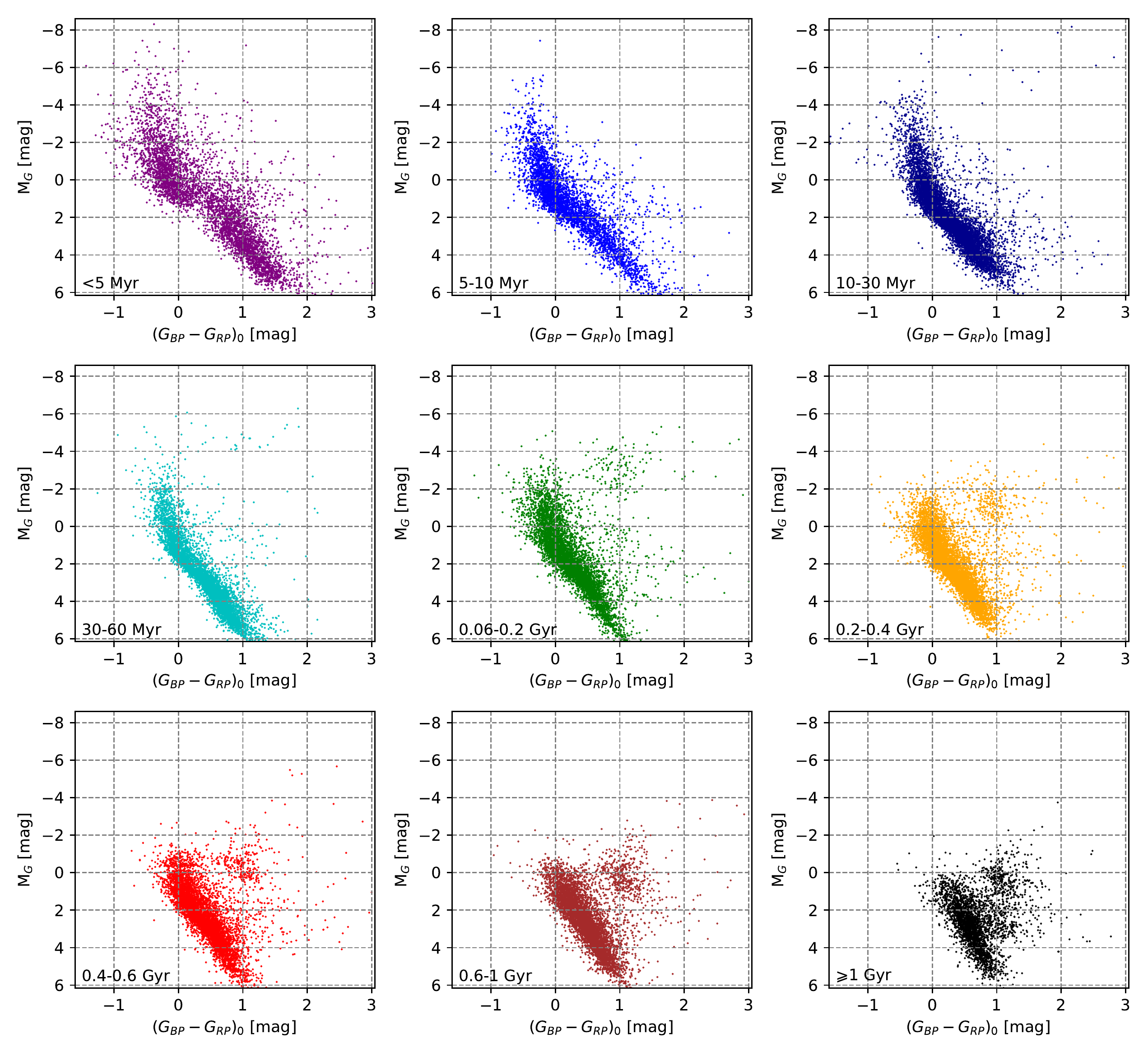}
	\caption{Hertzprung-Russell diagram for the  48,714 member stars in 1,656 new cluster candidates, which are divided into nine age intervals.}
	\label{hrds}
\end{center}
\end{figure*}

\begin{longrotatetable}
\begin{deluxetable*}{ccccccccccccccccccc}
	\tablecaption{Derived astrophysical parameters for the 1656 SC candidates found in this work. The positions, parallax, proper motions, and radial velocities of each cluster are calculated as the median value, and the dispersion of each astrometric value are also presented. The typical fitted accuracy for log(Age/yr), A$_0$, and distance modulus are 0.05 dex, 0.05~mag, and 0.05, respectively.}
	\label{tab_all}
	\tablewidth{1.0pt}
	\tabletypesize{\scriptsize}
\tablehead{CWNU ID  & l & b & $\sigma_{l}$ & $\sigma_{b}$ & n & parallax & $\sigma_{parallax}$ & pmra & $\sigma_{pmra}$ & pmdec & $\sigma_{pmdec}$ &RV &$\sigma_{RV}$ & N$_{RV}$ & log(Age/yr)& A$_0$ & m-M & class \\ 
 & [$^{\circ}$] & [$^{\circ}$] & [$^{\circ}$] & [$^{\circ}$] &  & [mas] & [mas] & [mas~  yr$^{-1}$] & [mas~  yr$^{-1}$] & [mas~  yr$^{-1}$] & [mas~  yr$^{-1}$] & [km~  s$^{-1}$] &[km~  s$^{-1}$] &  & [dex]  & [mag] &  & }
\startdata
\input{t1.tex}
\enddata
\end{deluxetable*}
\end{longrotatetable}
Figure~\ref{hrds} shows CMDs of all new cluster members in different age bins, where distance modulus, extinction, and reddening corrections have been applied to each member star. 
In this process, a polynomial function~\footnote{The extinction coefficient $k=c_1+c_2 \ast bp\_rp_0+c_3 \ast bp\_rp_0^2+c_4  \ast bp\_rp_0^3+c_5  \ast A_0+c6  \ast A_0^2+c_7  \ast A_0^3+c_8  \ast bp\_rp_0  \ast A_0+ c_9  \ast A_0  \ast bp\_rp_0^2+ c_{10}  \ast bp\_rp_0  \ast A_0^2$, here $c_1$ to $c_{10}$ values were adopted from the public auxiliary data provided by ESA/Gaia/DPAC/CU5 and prepared by Carine Babusiaux.} was used to compute the extinction coefficients~\citep{Danielski18,gaia_cmd}.
As can be seen from the figure, the overall CMDs are consistent with the characteristics of the SCs. As the age of a cluster increased, the length of the main sequence of the cluster gradually shortened, and the positions of the turn-off point and the red giant branch (RGB) stars were gradually shifted to dimmer magnitudes. For example, for clusters with ages of 30-60~Myr, 0.06-0.2~Gyr, and 0.2-0.4~Gyr, the RGB stars were located at M$_G~\sim$ -4~mag, -3~mag, and~-1~mag, respectively. Meanwhile, in each age interval, a small number of member stars deviated to the right of the main sequence, which may be due to inhomogeneous heavy reddening and/or field star contamination.

For the low absolute magnitude (2-6~mag) member stars in young clusters less than about 30 Myr, we noticed that some of them were low mass (about 0.5-2.5~M$_\odot$) stars and still in the pre-main sequence phase, which had slightly lower temperatures before they enter the main sequence~\citep{Henyey55}. For some old clusters, variables and stragglers may separate from the ideal tracks of evolved stars. Besides, binaries, inhomogeneous extinctions, multiple population, and/or stellar rotation~\citep[e.g.][]{Bastian09,Goudfrooij09,Licy19} also contributed to the extending of the observed evolution sequence. We have also provided Gaia EDR3 astrometry and photometry for all member stars, where some examples are shown in Table~\ref{table_mem}. The complete data set is available in the electronic version.

\begin{longrotatetable}
\begin{deluxetable*}{cccccccccccccccc}
	\tablecaption{The astrometric and photometric parameters of the member stars from Gaia EDR3~\citep{gaia2021}. The N$_{Re}$(from 1 to 11) is the number of responses frequency derived from DBSCAN (same to H22b).}
	\tablewidth{0pt}
	\label{table_mem}
	\tabletypesize{\scriptsize}
\tablehead{ source$\_$id&	l	&b	& ruwe &parallax&	parallax$\_$error	&pmra	&pmra$\_$error&	pmdec&	pmdec$\_$error& radia$\_$velocity &	radial$\_$velocity$\_$error &phot$\_$g$\_$mean$\_$mag	&bp$\_$rp  & N$_{Re}$& CWNU ID\\ 
   & [$^{\circ}$]  & [$^{\circ}$]  & & [mas] & [mas] & [mas~yr$^{-1}$] & [mas~yr$^{-1}$] & [mas~yr$^{-1}$] & [mas~yr$^{-1}$]   & [km~  s$^{-1}$] &[km~  s$^{-1}$] & [mag]&[mag]&  & }
	\startdata
\input{t2.tex}
	\enddata
\end{deluxetable*}
\end{longrotatetable}

We have cross-matched the new clusters with the stellar structures in ~\citet[][hereafter K20]{Kounkel20},  K20 identified $\sim$8,000 large-scale Galactic stellar structures that most of them span more than 100~pc, some candidates that appear in K20 were detected for the first time. Here we cross-matched all the member stars in this work with the member stars of K20. We found 21 ones may related to the new findings in this work, that means, more than 10\% of member stars are cross-matched (see `fraction' in Table~\ref{k20}). A more detailed research for the concentrated parts in K20 structures are planned to do in the near future.
\begin{deluxetable}{ccccccc}
\caption{The matched clusters found in this work and in K20.}
\label{k20}
\tablewidth{0.5pt}
\tablehead{\colhead{CWNU ID}& \colhead{n$_{CWNU}$} & \colhead{fraction$_{CWNU}$}&\colhead{Theia ID} & \colhead{n$_{Theia}$} & \colhead{fraction$_{Theia}$} & \colhead{n\_matched\_member}}
\startdata
1489 & 31 & 0.194 & 3385 & 42 & 0.143 & 6\\
1543 & 29 & 0.379 & 1750 & 115 & 0.096 & 11\\
1550 & 29 & 0.345 & 4212 & 72 & 0.139 & 10\\
1615  & 27 & 0.63 & 2773 & 123 & 0.138 & 17\\
1721  & 23 & 0.435 & 1973 & 80 & 0.125 & 10\\
1778 & 27 & 0.704 & 1769 & 131 & 0.145 & 19\\
1790 & 22 & 0.273 & 3841 & 43 & 0.14 & 6\\
1857 & 20 & 0.65 & 5205 & 128 & 0.102 & 13\\
1906 & 19 & 0.474 & 2871 & 89 & 0.101 & 9\\
2314 & 66 & 0.439 & 1779 & 148 & 0.196 & 29\\
2458 & 42 & 0.476 & 2665 & 61 & 0.328 & 20\\
2535 & 106 & 0.189 & 2133 & 104 & 0.192 & 20\\
2574 & 73 & 0.233 & 2453 & 167 & 0.102 & 17\\
2576 & 73 & 0.247 & 3590 & 53 & 0.34 & 18\\
2656 & 57 & 0.316 & 3340 & 132 & 0.136 & 18\\
2694 & 52 & 0.481 & 3455 & 194 & 0.129 & 25 \\
2698 & 52 & 0.135 & 3404 & 40 & 0.175 & 7\\
2711  & 50 & 0.22 & 2069 & 99 & 0.111 & 11\\
2722 & 49 & 0.347 & 1905 & 174 & 0.098 & 17\\
2723 & 48 & 0.312 & 3163 & 83 & 0.181 & 15\\
2760 & 45 & 0.2 & 3605 & 87 & 0.103 & 9\\
\enddata
\end{deluxetable}

\section{Analysis}\label{sec:analysis}

\subsection{ Classifications and examples}\label{sec:classification}
\begin{figure*}
\begin{center}	
			\includegraphics[width=0.9\linewidth]{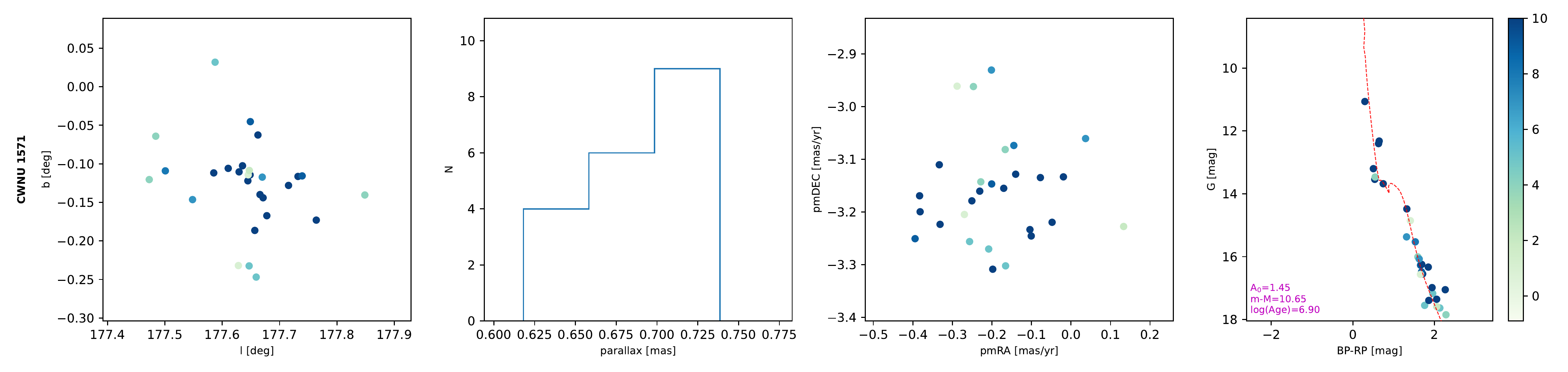}	
			\includegraphics[width=0.9\linewidth]{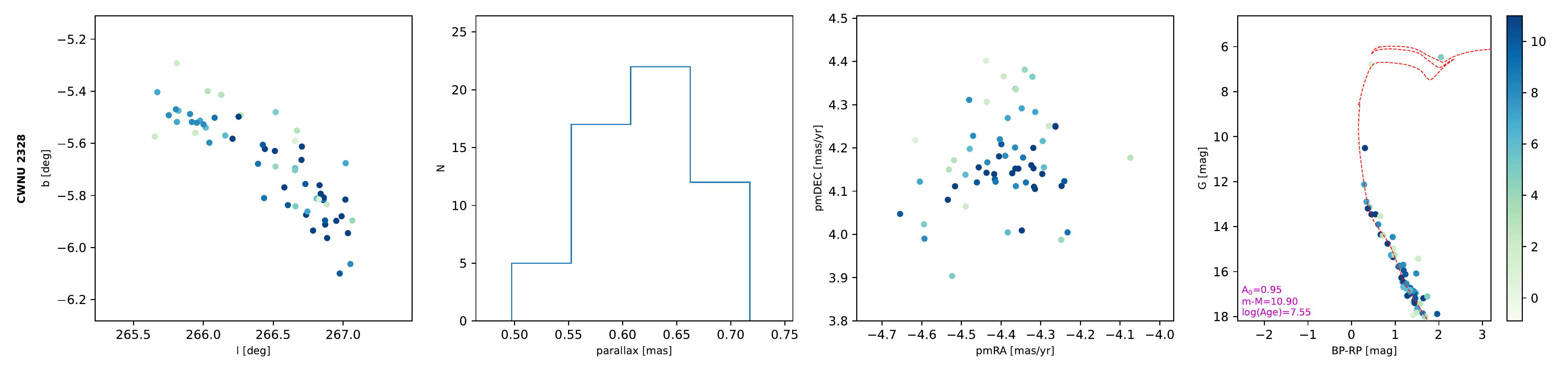}
			\includegraphics[width=0.9\linewidth]{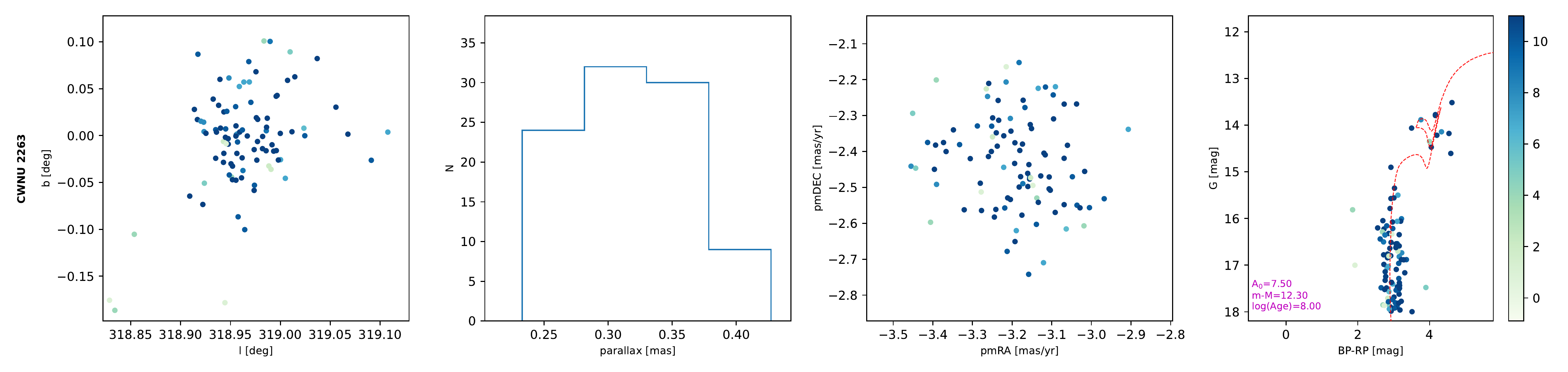}
				\includegraphics[width=0.9\linewidth]{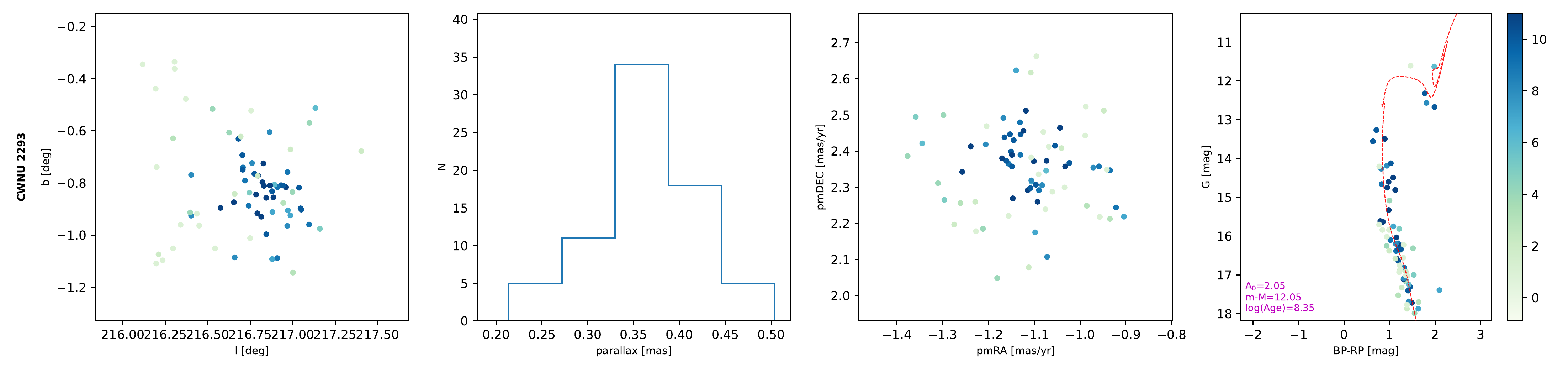}
					\includegraphics[width=0.9\linewidth]{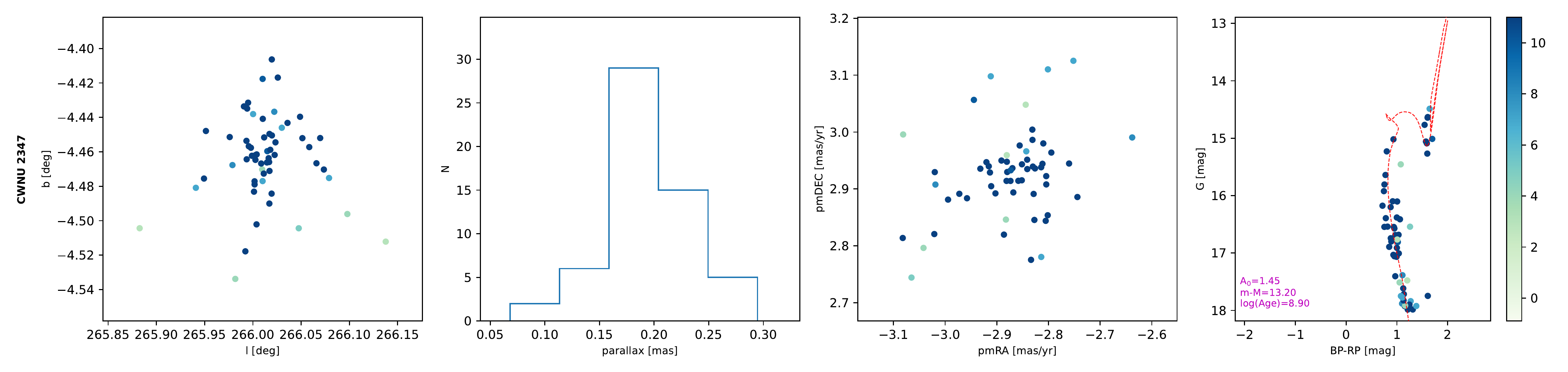}
					\includegraphics[width=0.9\linewidth]{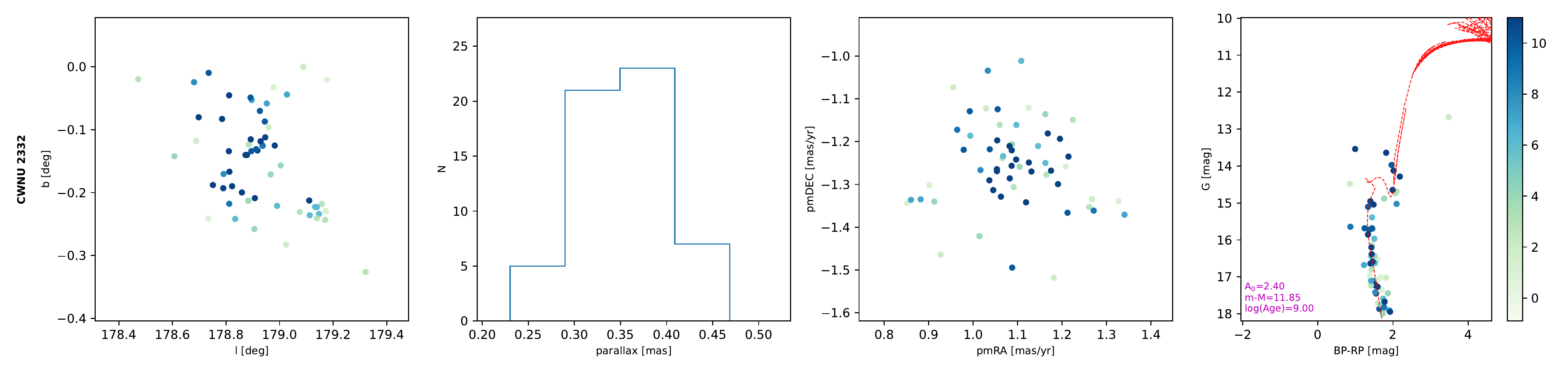}
	\caption{Examples of the new candidates in class 1. The four subplots show the spatial distributions, parallax statistics, proper motion distributions, and observed CMDs of the cluster members. The best-fitting isochrones are also shown in the figure. The color bars represent the frequency (N$_{Re}$) with which a given star is found by DBSCAN under different the $k$ values.}
	\label{cls1}
\end{center}
\end{figure*}

\begin{figure*}
\begin{center}
	\includegraphics[width=0.9\linewidth]{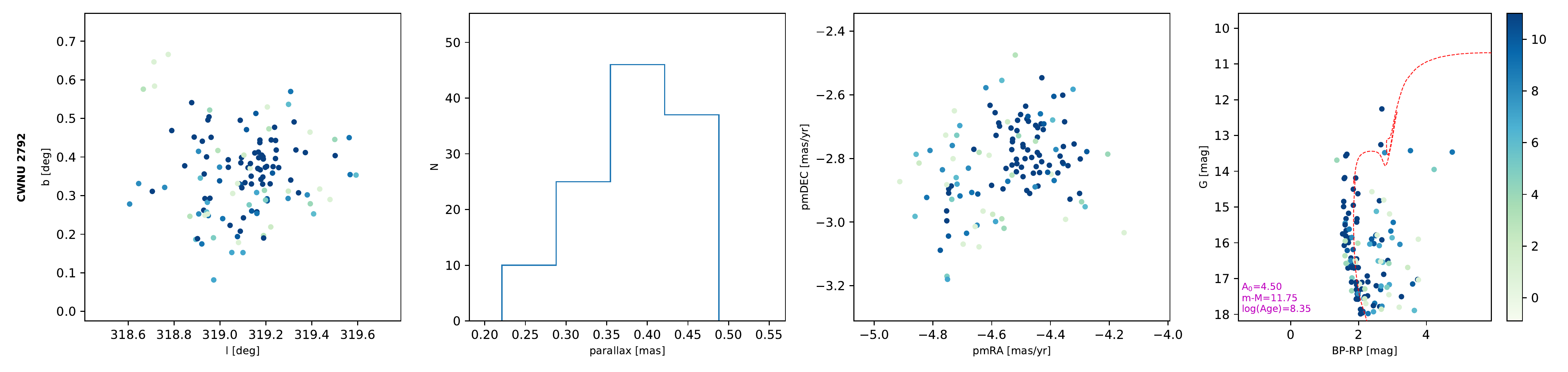}
	\includegraphics[width=0.9\linewidth]{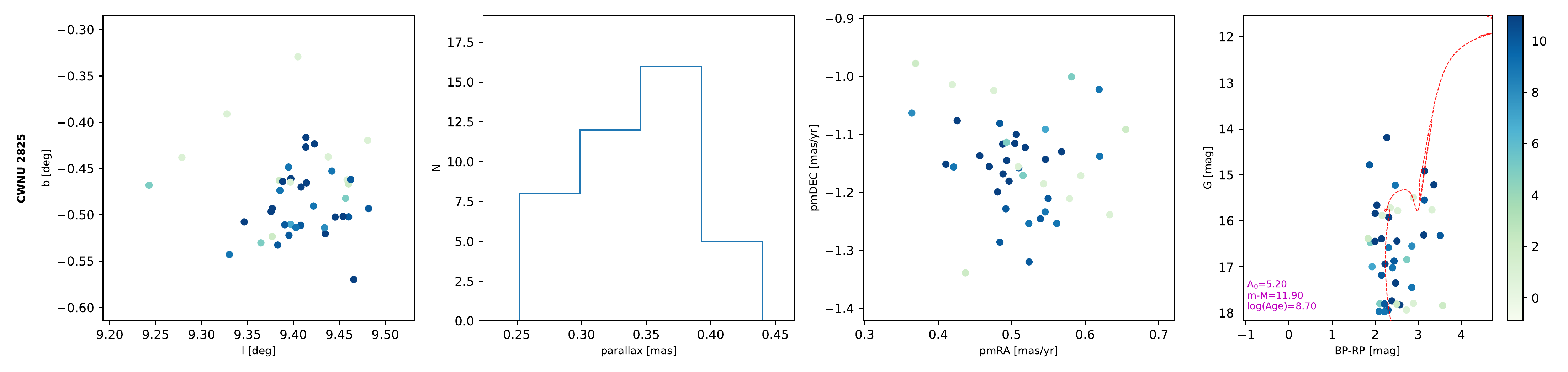}
	\caption{Same as Fig.~\ref{cls1}, but for the new candidates in class 2.}
	\label{cls2}
\end{center}
\end{figure*}

\begin{figure*}
\begin{center}
\includegraphics[width=0.9\linewidth]{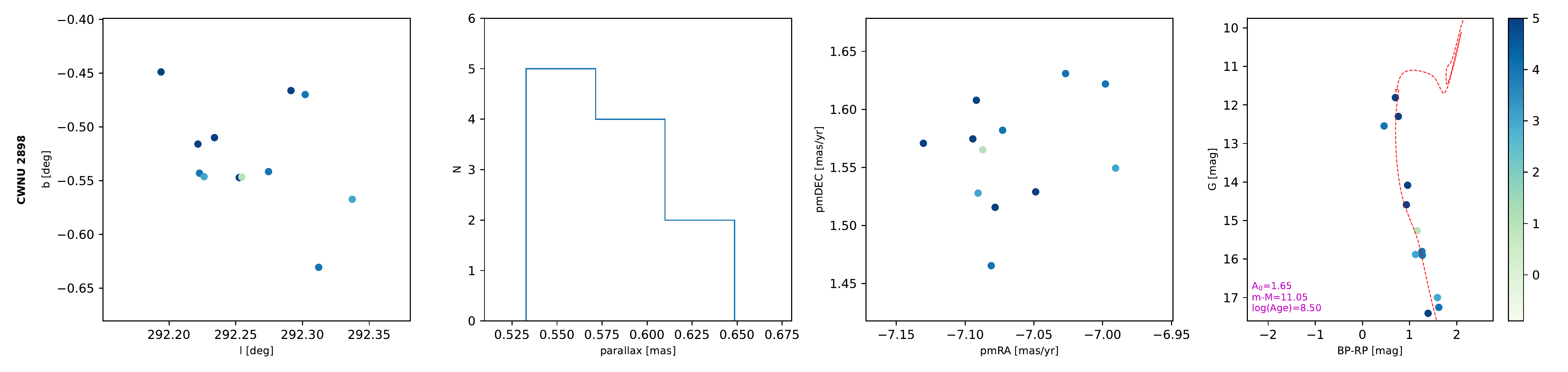}
	\includegraphics[width=0.9\linewidth]{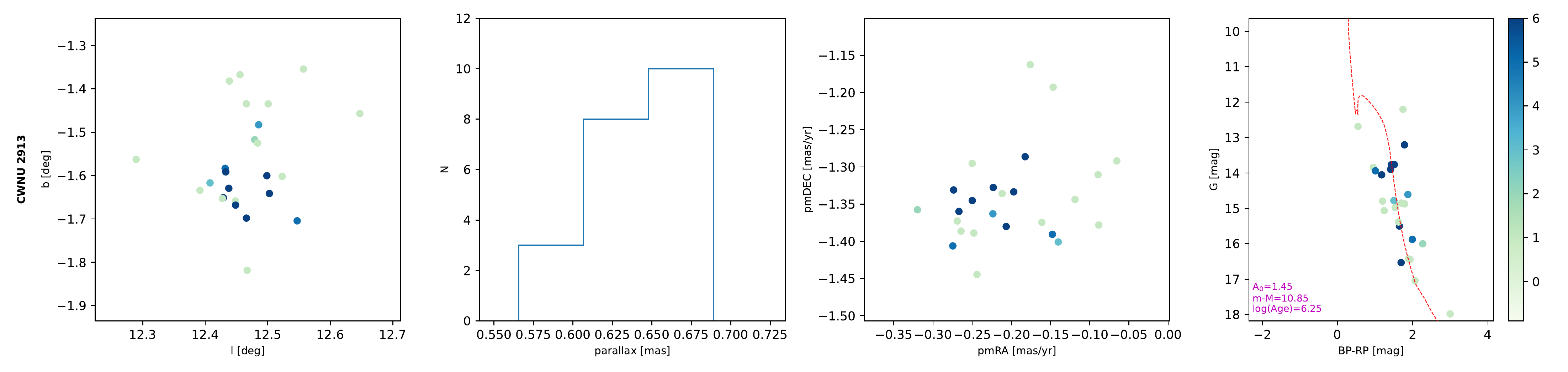}
	\caption{Same as Fig.~\ref{cls1}, but for the new candidates in class 3.}
	\label{cls3}
\end{center}
\end{figure*}

Similar to H22a,b, we divided the new candidates into three classes. After visual inspection of all candidates, we classified  1,491 cluster candidates with clear main sequences in the CMD into the first class (Fig.~\ref{cls1}), and their identities were labeled CWNU 1271 to CWNU~2761. As shown in Fig.~\ref{cls2}, 110 candidates with more reddened member stars in the CMDs were classified into the second class, and their identifications were marked as CWNU~2762 to CWNU~2871. For 55 candidates whose main sequences were too short and had obvious abnormal gaps, they were classified into the third class (Fig.~\ref{cls3}) and labelled as CWNU~2872 to CWNU~2926. The mean number of member stars for class 1/2/3 candidates are 30, 28, and 14, respectively. Full figures of the new candidates of classes 1-3 can be viewed in Figure sets 1-3, respectively.


\begin{figure*}
\begin{center}	
	\includegraphics[width=0.9\linewidth]{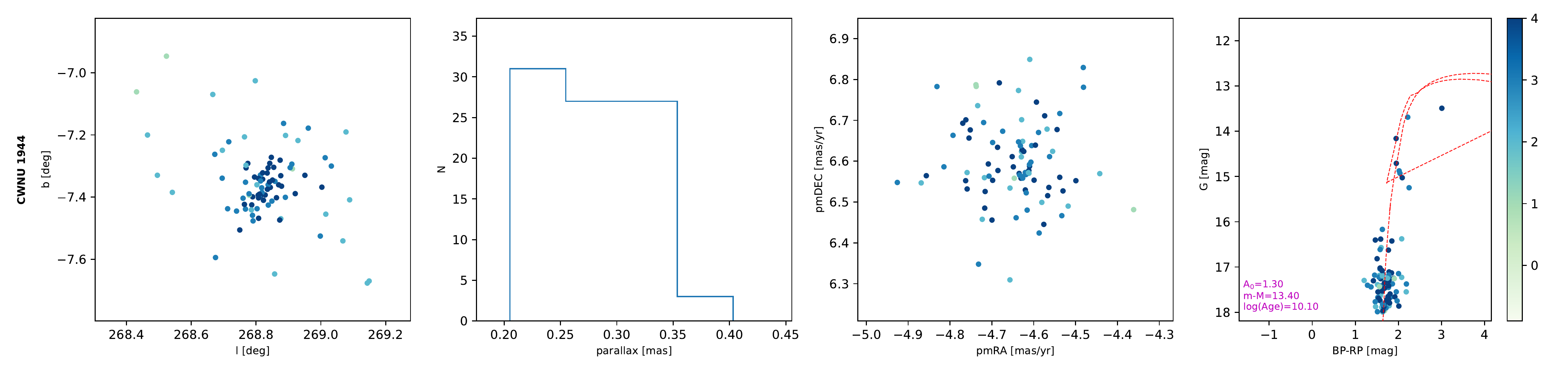}
	\caption{Same as Fig.~\ref{cls1}, but for the new globular cluster candidate CWNU~1944.}
	\label{glo}
\end{center}
\end{figure*}

Most clusters have the photometric signature of an OC, but we also found a candidate cluster that resembles a globular cluster, labeled CWNU~1944 (Fig.~\ref{glo}). This cluster is located only ~500~pc from the Galactic plane and about 3.5~kpc from the Solar System, and it only contains evolved stars below G = 18~mag. It had not been detected in past globular cluster searches, possibly because it is located very close to the Galactic plane and not in the direction of the Milky Way bulge.

\subsection{Radial velocity}\label{sec:rv}    
\begin{figure*}
\begin{center}
	\includegraphics[width=0.5\linewidth]{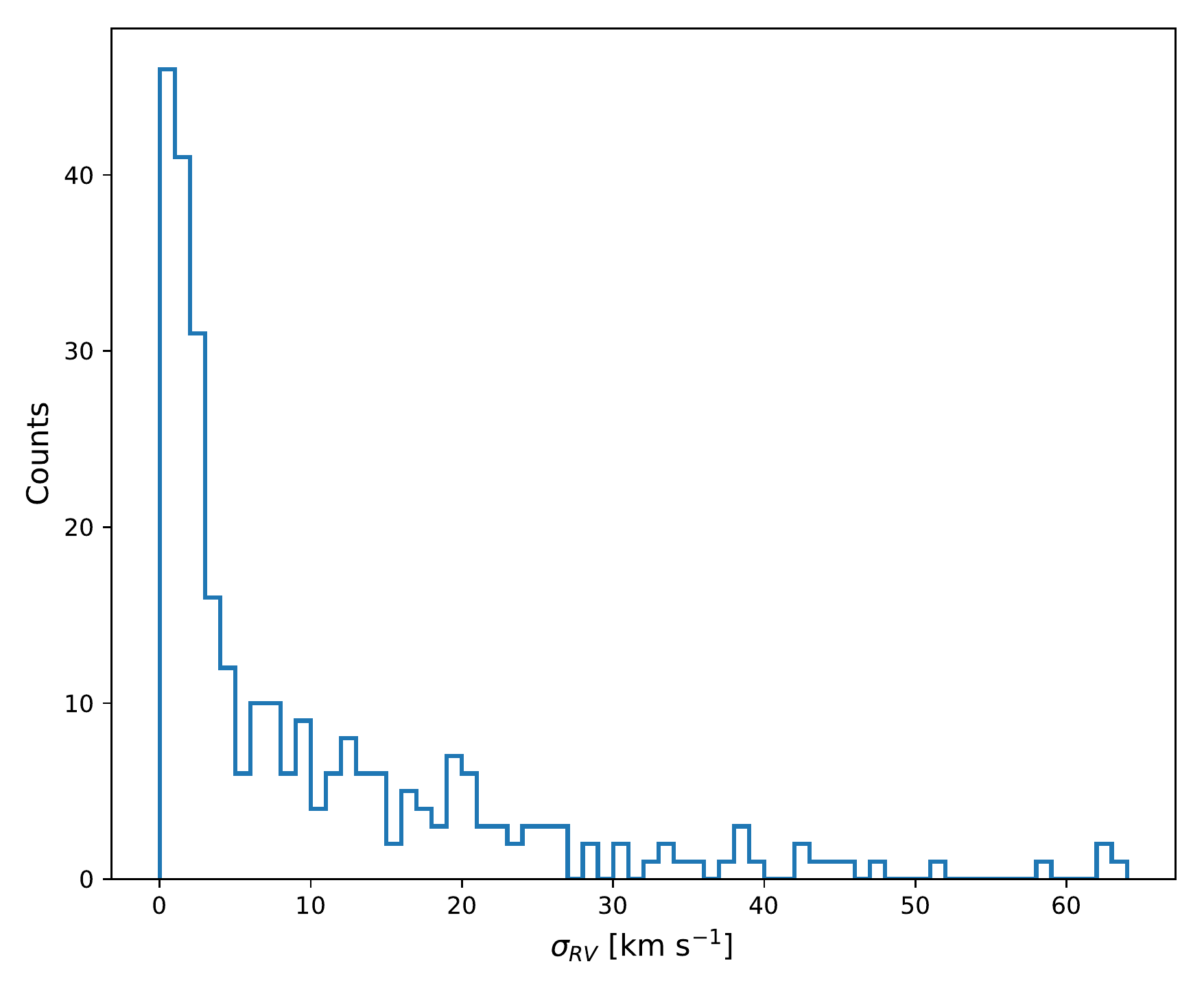}
	\caption{Histogram of the RV dispersion for 328 new cluster candidates with more than one matched member star in Gaia DR3 RV database.}
	\label{rv}
\end{center}
\end{figure*}

During our preparation for this work, the latest Gaia Data Release~\citep[DR3,][]{gaiadr3} published radial velocity (RV) measurements for more than 30 million stars, and we selected stars with the uncertainties of less than 5 \kms to cross-match with all member stars of the 1,656 new cluster candidates.
Of the 426 clusters, only one star was matched.
In addition, there are 168 candidates with 2 matched stars, of which 146 are in class~1, the median value of RV dispersion (MRVD) is 7.7~\kms; 21 of them are in class~2, the MRVD is 14.3~\kms; and only 1 is in class~3, its RV dispersion is 0.1~\kms.
There are 140 new candidates in class~1 possess more than two matched member stars, and the MRVD is 3.0~\kms; 17 of them are in class 2, the MRVD is 7.8~\kms; of which 3 are in class~3, and the MRVD is 0.8~\kms.
The histogram of the radial velocity dispersion of all 328 SC candidates with more than one matched member star is shown in Fig.~\ref{rv}. It can be inferred that field star contamination still exists in the results, which is consistent with the conclusion obtained by ~\citet{castro22}.

\subsection{Parallax and comparison with Gaia DR2 clusters}\label{sec:plx}   
\begin{figure*}
\begin{center}
	\includegraphics[width=0.9\linewidth]{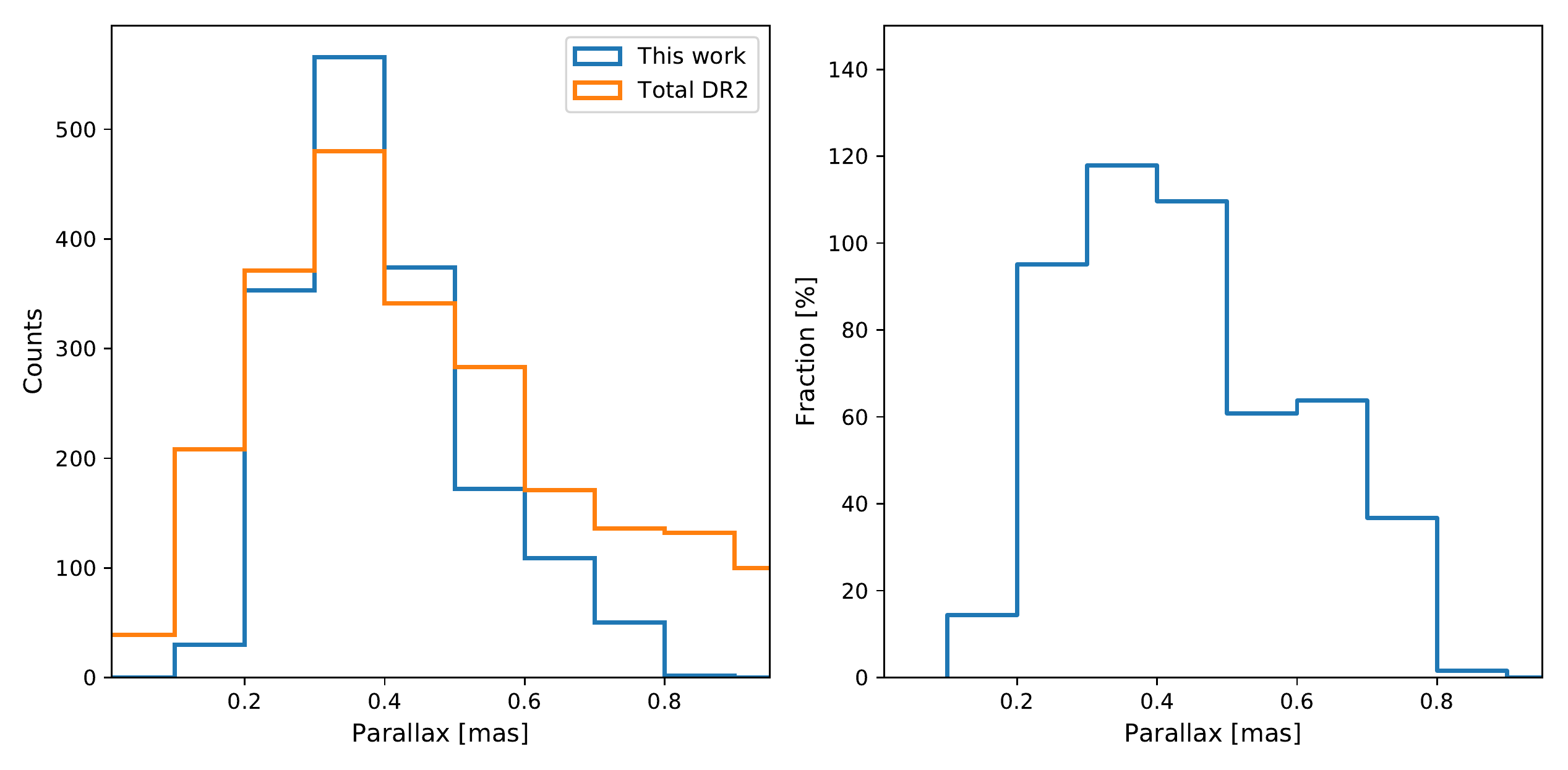}
	\caption{The left panel shows a histogram of the parallaxes. Orange: cluster sample of Gaia DR2 OCs.
Blue: The sample found in this work. The right panel shows the proportion of increase in the number of Gaia DR2 OCs at the different parallaxes.}
	\label{plx}
\end{center}
\end{figure*}

Compared to the cluster sample in Gaia DR2~\citep[][in total of 2,793 clusters]{CG20,Ferreira19,Ferreira20,ferreira21,Qin20,he21,he22a,Hunt21,Casado21}, the number of new results present in this work is equivalent to ~60\% of the population of Gaia DR2 open clusters. The histograms in Fig.~\ref{plx} show a comparison of the cluster parallaxes in this work with those in Gaia DR2. The parallax values of the new clusters are mostly above 0.2~mas. Especially around 0.4~mas, the population of SCs increased by more than 120\%. Recently, a large number of new clusters have also been discovered in Gaia EDR3, and our results have more than doubled the total number of Gaia EDR3 clusters. This showed that the improved accuracy of Gaia's astrometric parameters have great advantages for SC searches. At the same time, this also reflected that our method had high efficiency for the search of SCs within 5~kpc of the Solar System.

\subsection{Age and extinction}\label{sec:Age} 
\begin{figure*}
\begin{center}
	\includegraphics[width=0.45\linewidth]{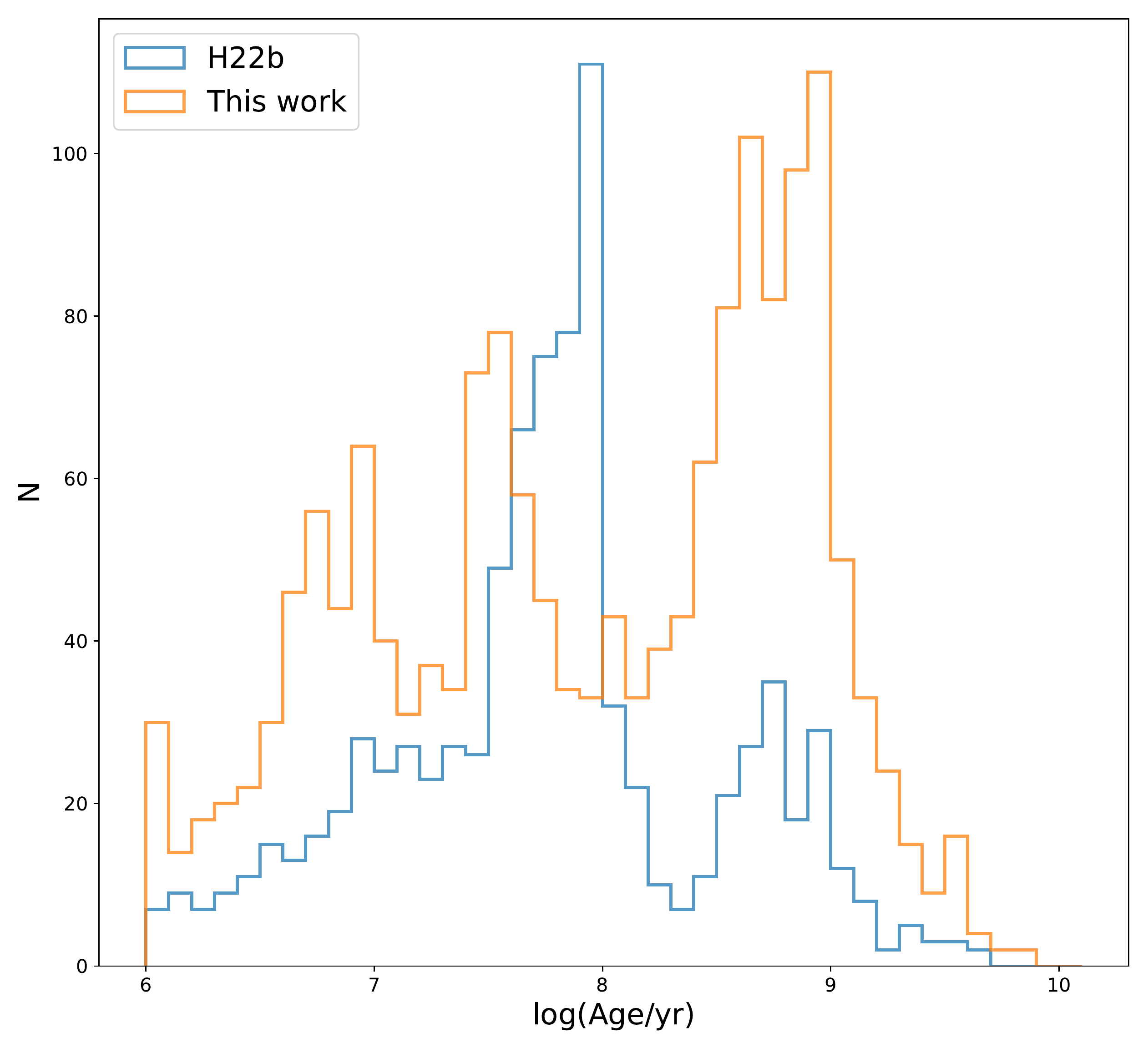}
		\includegraphics[width=0.45\linewidth]{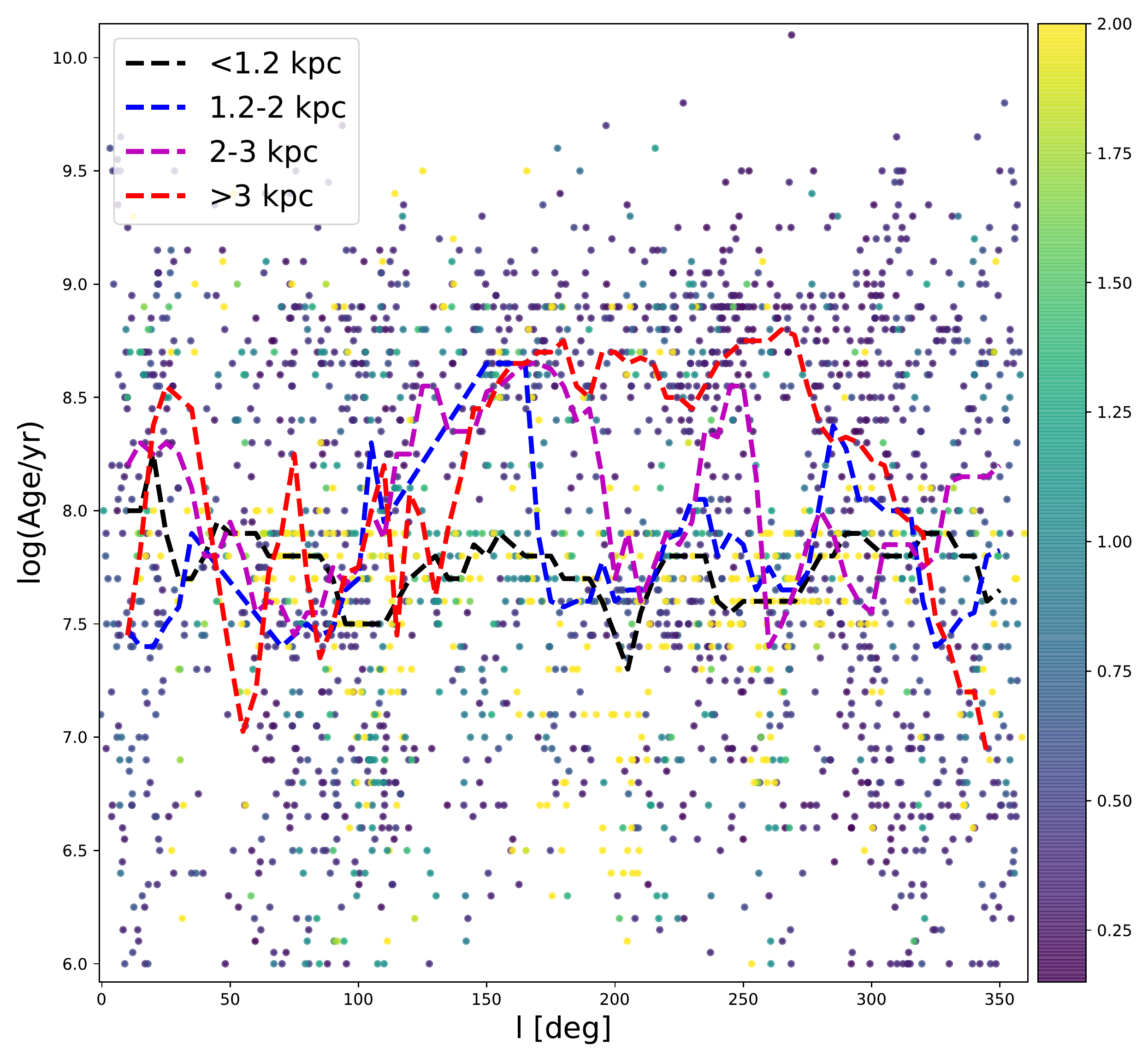}
	\caption{Left: Histograms of the cluster ages in H22b and in this work. Right: Galactic longitude-ages (only SC candidates in this work and in H22b) distributions, color-coded by parallax, where the colored dash lines show the median ages at the different longitudes.}
	\label{age}
\end{center}
\end{figure*}

In Figure~\ref{age}, we have displayed the age distributions of the SC candidates in the different catalogs. It is obvious that the new SC candidates are younger than those within 1.2~kpc of the Solar System. This may be due to the large number of star-forming regions in the solar neighborhood, such as Taurus, Orion, Vela, etc. (H22b). For distant star-forming regions, dust can significantly obscure the SCs behind/in them. The right panel shows the ages of the new candidates as a function of Galactic longitude and parallax. It can be seen that in the third quadrant, most of the new candidates with distances beyond 2~kpc are old clusters, and there is no obvious difference in other directions.

\begin{figure*}
\begin{center}
\includegraphics[width=0.45\linewidth]{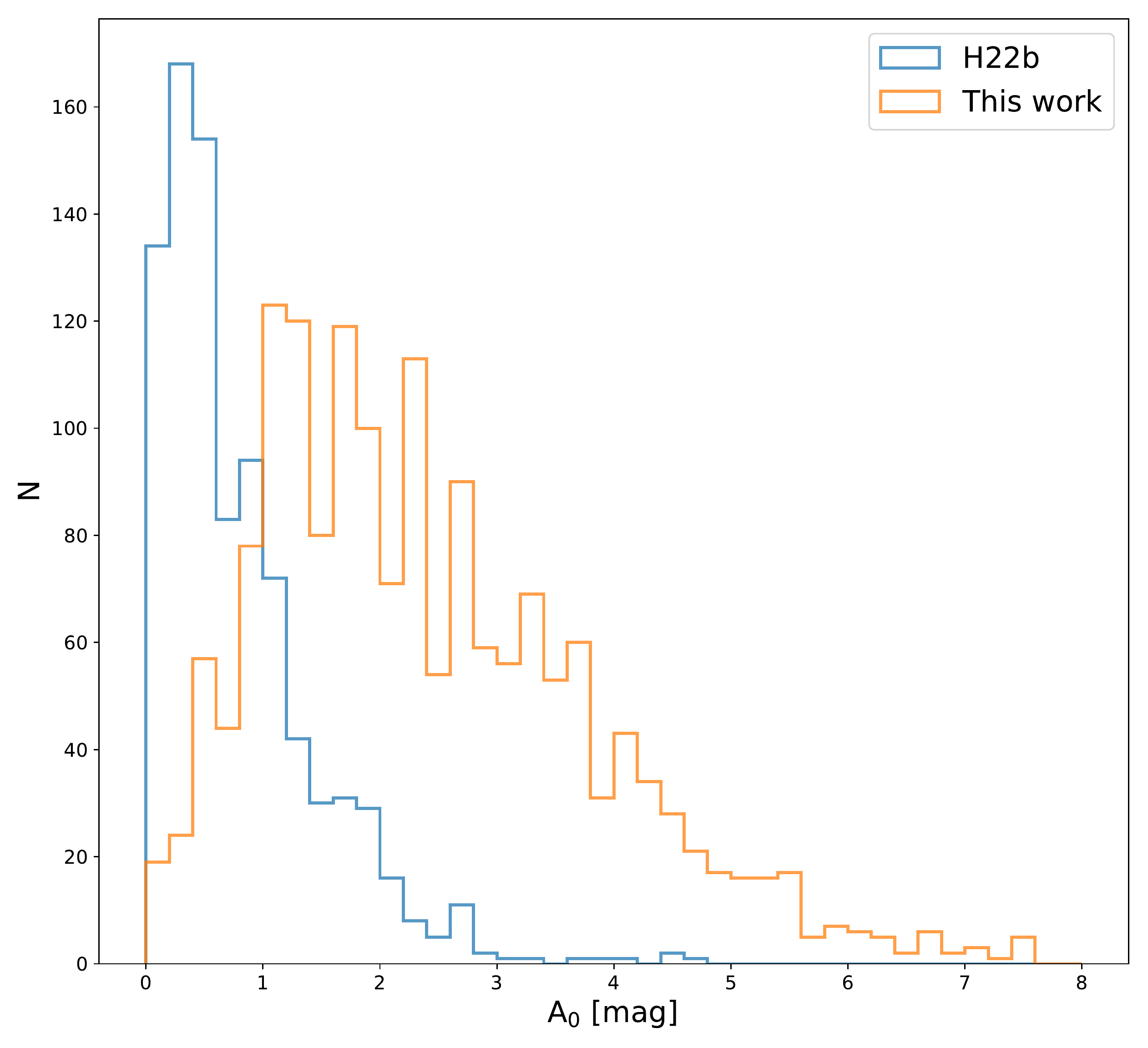}
	\includegraphics[width=0.45\linewidth]{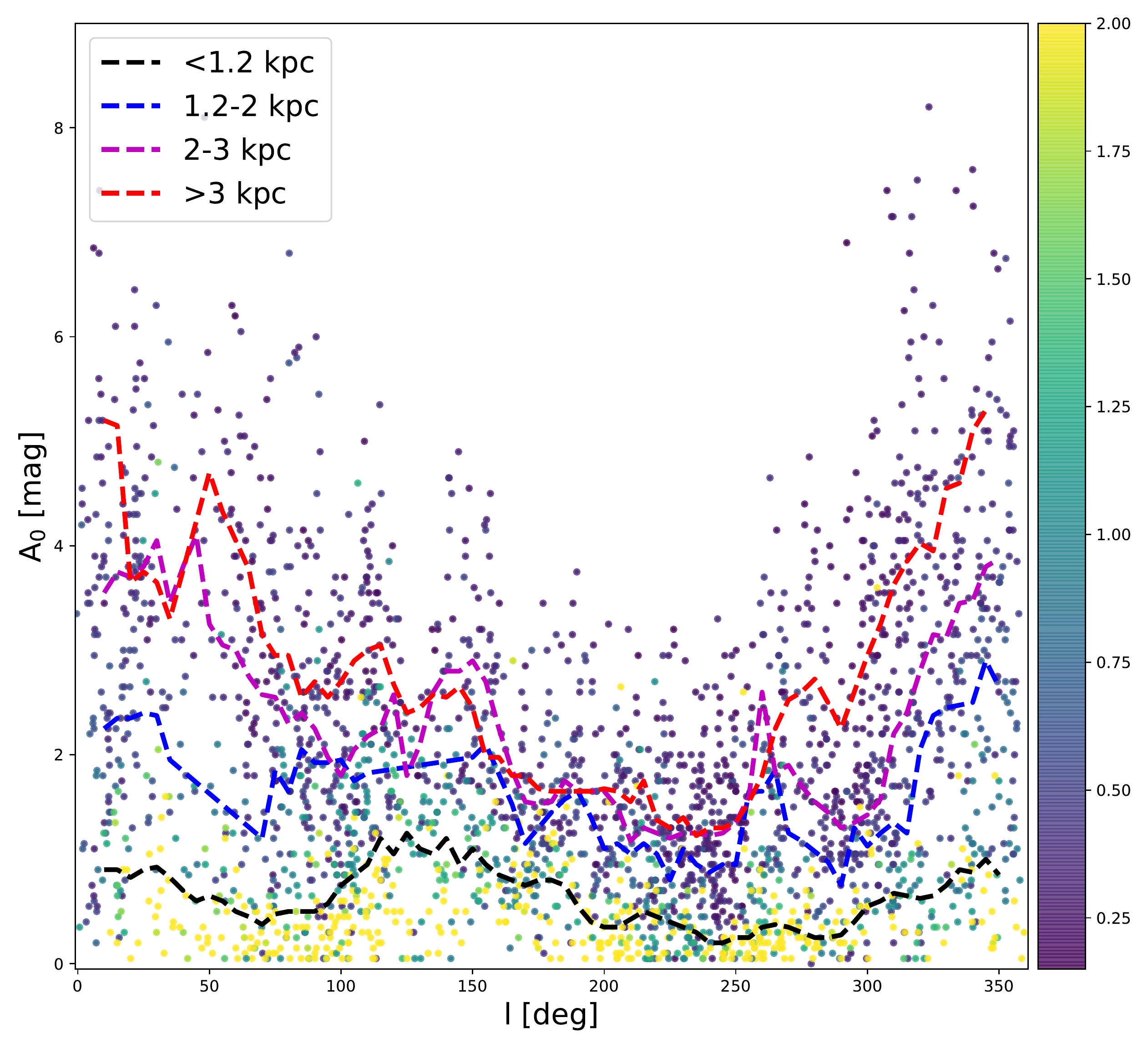}
	\caption{Same as the Fig.~\ref{age} but for the line-of-sight extinction.}
	\label{ag}
\end{center}
\end{figure*}

The extinction distributions are shown in Fig.~\ref{ag}. The new candidates have extinction values below 8~mag, and most of them are around 2~mag, which are undoubtedly more extinguished than clusters in the solar vicinity. The right panel shows the extinction values of the new candidates as a function of Galactic longitude and parallax. In most directions, the dust extinction value become significantly larger with increasing distance, except in the third quadrant of the Milky Way, which also have the smallest extinctions. It can be inferred that, in the regions outside this interval, dust and molecular gas appear to be absent, which may be a direct reason for the lack of young SCs therein (Fig.~\ref{age}). However, these statistical analyses were only showed the results obtained for our new samples, and the homogeneous isochrone fits are required to obtain more unbiased values of the extinction distribution.

\subsection{Proper motion dispersions and apparent radii}\label{sec:dispersion} 
\begin{figure*}
\begin{center}
	\includegraphics[width=0.95\linewidth]{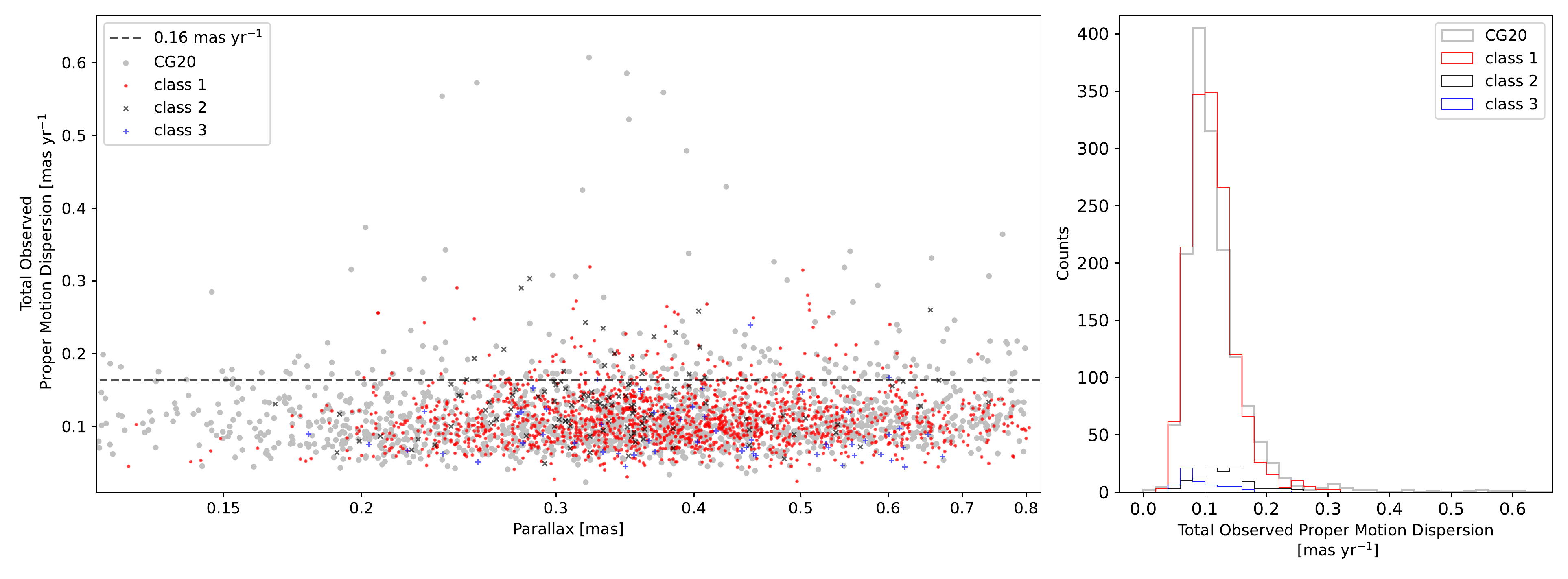}
\includegraphics[width=0.95\linewidth]{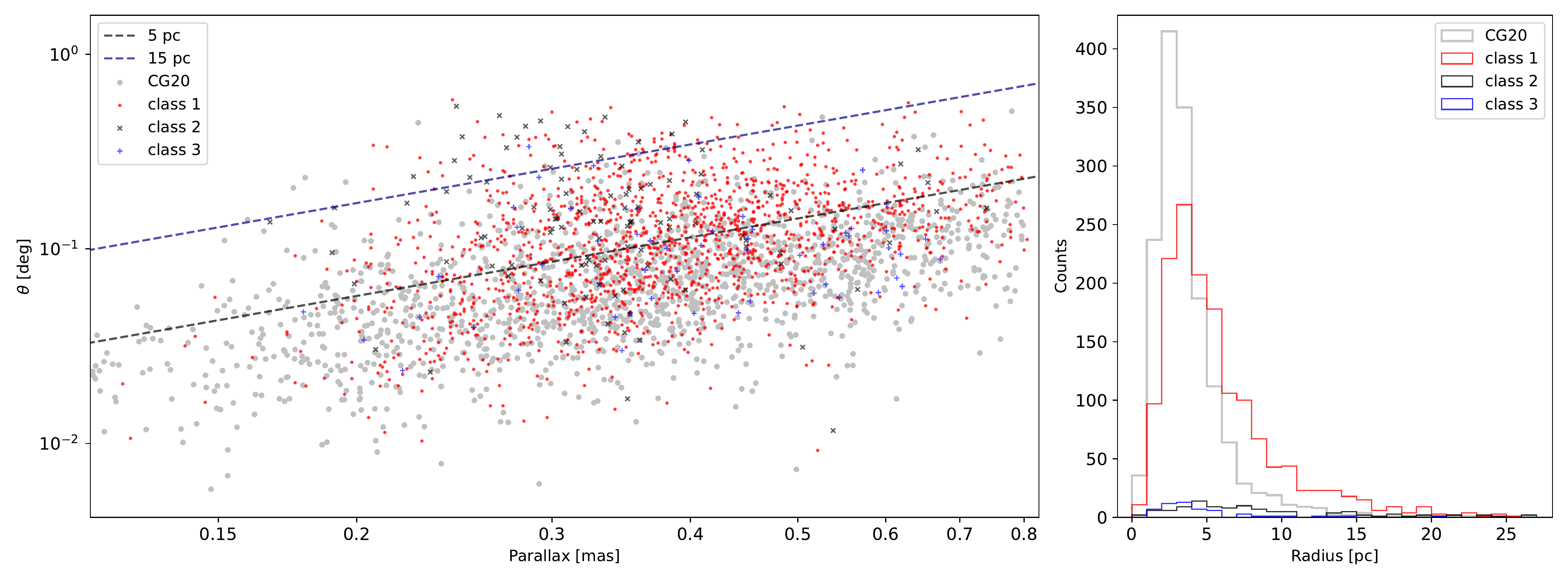}
	\caption{Upper panels: comparison of the total observed proper motion dispersions from CG20 (member stars were cross-matched in Gaia EDR3) and this work, the x-axis in the left plot presents a logarithmic scale parallax, and the right plot shows the histograms. Bottom panels: same as the upper panels but for the apparent radius, the black and blue dashed lines indicate the angular size corresponding to 5 and 15 pc, respectively.}
	\label{pmdisper}
\end{center}
\end{figure*}

The proper motion dispersions of cluster members are necessary to measure whether an SC candidate is a true cluster, which has been pointed out and verified by~\citet{CG20_0} and ~\citet{dias21}. In Gaia EDR3, although the accuracies of the proper motions improved by more than three times relative to those in Gaia DR2, they were still not enough to resolve linear velocity dispersions below 1~km s$^{-1}$ for clusters beyond 1.2~kpc. However, the dispersion values of the proper motions could also be used to distinguish whether stars were kinematically linked or not. As shown in Fig.~\ref{pmdisper}, the total proper motion dispersion distributions of the new candidates are consistent with the values in CG20. Specifically, 90\% of the candidates had a total dispersion below 0.16~mas~yr$^{-1}$. In addition, the sizes of the apparent radii ~\citep[estimated as $\theta = \sqrt{\sigma_{l}^2 \cdot cos^2(b)+\sigma_{b}^2}$,][]{castro22,he22b} of the clusters in the two catalogs were also similar, which mostly smaller than 5 to 15~pc. From the comparison of the above two aspects, combined with the CMD characteristics of the new candidates, it can be seen that the new candidates are clusters with the characteristics of genuine SCs.

\section{Summary}\label{sec:summary}
We performed a blind search of the Galactic plane beyond 1.2~kpc of the Solar System based on Gaia EDR3 data. We used the DBSCAN algorithm, which is widely used in cluster searches, and we modified the method and procedure for inputting data and computing clustering parameters. After clustering and cross-matching with existing cluster catalogs, we found 1,656 new SC candidates. We have listed the physical results of all discoveries, as well as member star information. Photometric data of these candidates present the appearance of conventional cluster CMDs, and their sizes and proper motion dispersions are also consistent with currently identified clusters. 
It is reasonable to infer that these candidates are real SCs.

The new SCs were mostly located within 5~kpc of the Solar System, and the new findings presented in this work increased the cluster sample size by more than 30\%, expanding the total number of Galactic clusters to about 6,000. Nearly half of these clusters were discovered by our works, demonstrating the high efficiency of the method that we utilized. The new clusters were older than those near the Solar System and suffered from a significantly greater degree of extinction than nearby clusters. This meant that the current cluster search was still affected by extinction, and since fainter old clusters were difficult to detect, it is reasonable to believe that there are many undiscovered clusters may still hidden from Gaia's view. 

SCs are important laboratories for studying the evolution of stars and the clusters themselves, and are also good tracers for exploring the structure of the Galaxy. More abundant stellar radial velocity and physical information have been released in Gaia DR3~\citep{Blanco22}, which provided a great opportunity to study cluster membership and kinematics. At present, the sample of SCs is becoming more and more abundant, and in order to make the total known sample unbiased and complete, the published cluster catalogs need to be cross-matched with each other. At the same time, whether the new record clusters in different published catalogs are real SCs requires homogeneous member star identification and confirmation, which will be a challenging but necessary work in the near future. 

\section{Acknowledgements}
We thank the referee for the constructive comments and suggestions that helped to improve this paper. This work has made use of data from the European Space Agency (ESA) mission GAIA (\url{https://www.cosmos.esa.int/gaia}), processed by the GAIA Data Processing and Analysis Consortium (DPAC,\url{https://www.cosmos.esa.int/web/gaia/dpac/consortium}). Funding for the DPAC has been provided by national institutions, in particular the institutions participating in the GAIA Multilateral Agreement.
This work is supported by Fundamental Research Funds of China West Normal University (CWNU, No.21E030), the Innewtion Team Funds of CWNU, and the Sichuan Youth Science and Technology innovation Research Team (21CXTD0038), K.W. is supported by the Sichuan Science and Technology Program (No.2020YFSY0034), the National Natural Science Foundation of China (NSFC, No.12003022). 
L.Y.P is supported by the National Key Basic R\&D Program of China via 2021YFA1600401, the NSFC under grant 12173028, and the Chinese Space Station Telescope project with NO. CMS-CSST-2021-A10.

\bibliographystyle{aasjournal} 
\bibliography{noc} 

\end{CJK*}

\end{document}

%% file: t1.tex
1271 & 115.223 & -6.073 & 0.095 & 0.083 & 45 & 0.44 & 0.05 & -5.10 & 0.09 & -1.58 & 0.07& -47.99 &  &  1 & 8.15 & 0.80 & 11.65&  1 \\ 
1272 & 276.428 & 1.151 & 0.088 & 0.055 & 26 & 0.39 & 0.04 & -5.35 & 0.05 & 2.85 & 0.08& 12.51 &  &  1 & 8.15 & 2.50 & 11.80&  1 \\ 
1273 & 237.917 & 0.226 & 0.108 & 0.067 & 45 & 0.36 & 0.03 & -3.23 & 0.08 & 3.15 & 0.08&   &  &  0 & 7.55 & 1.25 & 11.95&  1 \\ 
1274 & 195.524 & 1.377 & 0.060 & 0.072 & 45 & 0.33 & 0.06 & 0.12 & 0.08 & -1.00 & 0.11& 34.73 & 1.42&  2 & 8.90 & 1.30 & 12.10&  1 \\ 
1275 & 319.505 & -1.270 & 0.045 & 0.029 & 45 & 0.35 & 0.07 & -3.76 & 0.08 & -2.86 & 0.10&   &  &  0 & 6.65 & 5.60 & 12.15&  1 \\ 
1276 & 98.916 & -2.338 & 0.137 & 0.060 & 44 & 0.43 & 0.03 & -2.46 & 0.10 & -3.15 & 0.09& -58.31 & 8.99&  2 & 8.65 & 1.30 & 11.70&  1 \\ 
1277 & 229.720 & 1.855 & 0.056 & 0.064 & 25 & 0.29 & 0.07 & -1.66 & 0.06 & 0.86 & 0.06&   &  &  0 & 8.50 & 1.00 & 12.65&  1 \\ 
1278 & 149.583 & 0.977 & 0.128 & 0.059 & 44 & 0.30 & 0.03 & -0.00 & 0.10 & -1.22 & 0.07& -61.80 & 14.69&  2 & 9.00 & 1.75 & 12.25&  1 \\ 
1279 & 351.445 & -0.474 & 0.107 & 0.082 & 25 & 0.59 & 0.03 & 1.41 & 0.08 & -0.51 & 0.09&   &  &  0 & 7.05 & 3.80 & 11.20&  1 \\ 
1280 & 253.762 & 2.665 & 0.030 & 0.032 & 44 & 0.20 & 0.05 & -3.31 & 0.06 & 4.18 & 0.08& 72.04 & 2.07&  3 & 8.80 & 1.35 & 13.10&  1 \\ 
1281 & 207.682 & 0.179 & 0.132 & 0.194 & 44 & 0.64 & 0.04 & -1.97 & 0.08 & 0.82 & 0.07&   &  &  0 & 7.60 & 1.10 & 10.85&  1 \\ 
1282 & 354.381 & -1.454 & 0.037 & 0.023 & 25 & 0.30 & 0.04 & -1.00 & 0.11 & -2.50 & 0.09&   &  &  0 & 6.20 & 5.05 & 13.15&  1 \\ 
1283 & 320.305 & -1.431 & 0.099 & 0.073 & 25 & 0.36 & 0.04 & -4.76 & 0.10 & -3.77 & 0.08& -56.12 &  &  1 & 6.35 & 2.30 & 12.60&  1 \\ 
1284 & 9.397 & -0.549 & 0.029 & 0.024 & 44 & 0.37 & 0.04 & -1.25 & 0.08 & -2.59 & 0.06& -29.40 & 1.79&  5 & 8.50 & 4.60 & 12.20&  1 \\ 
1285 & 52.740 & 0.553 & 0.058 & 0.042 & 43 & 0.36 & 0.06 & -1.72 & 0.10 & -5.49 & 0.06& 20.10 &  &  1 & 8.20 & 4.35 & 11.95&  1 \\ 
... & ... & ... & ... & ... & ... & ... & ... & ... & ... & ... & ... & ... & ... & ...& ...& ... & ...& ...\\ 
2763 & 295.978 & 2.174 & 0.020 & 0.063 & 31 & 0.20 & 0.04 & -6.71 & 0.06 & 1.09 & 0.06&   &  &  0 & 6.80 & 1.55 & 13.10&  2 \\ 
2764 & 309.748 & 0.211 & 0.070 & 0.099 & 53 & 0.30 & 0.04 & -6.01 & 0.10 & -1.84 & 0.09& -61.28 &  &  1 & 7.10 & 7.15 & 12.40&  2 \\ 
2765 & 349.528 & -1.543 & 0.049 & 0.066 & 56 & 0.37 & 0.03 & -0.32 & 0.10 & -2.72 & 0.07& -37.17 & 6.04&  5 & 9.15 & 2.75 & 11.85&  2 \\ 
2766 & 324.124 & 2.452 & 0.197 & 0.045 & 17 & 0.35 & 0.04 & -4.62 & 0.09 & -3.52 & 0.08&   &  &  0 & 9.10 & 2.80 & 12.70&  2 \\ 
2767 & 317.700 & -0.951 & 0.038 & 0.016 & 16 & 0.34 & 0.02 & -3.37 & 0.10 & -2.91 & 0.08& -43.77 & 0.25&  2 & 8.25 & 6.45 & 12.15&  2 \\ 
2768 & 146.081 & 2.348 & 0.037 & 0.539 & 23 & 0.25 & 0.04 & 0.10 & 0.08 & -0.65 & 0.12& -38.13 & 9.02&  3 & 8.90 & 1.85 & 12.55&  2 \\ 
2769 & 316.844 & -0.378 & 0.034 & 0.015 & 18 & 0.35 & 0.03 & -5.02 & 0.04 & -2.70 & 0.04& -21.96 &  &  1 & 7.85 & 7.15 & 12.05&  2 \\ 
2770 & 226.380 & -2.781 & 0.059 & 0.124 & 25 & 0.17 & 0.04 & -0.35 & 0.08 & 0.99 & 0.10&   &  &  0 & 6.65 & 3.20 & 16.15&  2 \\ 
2771 & 349.976 & 3.586 & 0.047 & 0.077 & 17 & 0.32 & 0.04 & -1.46 & 0.07 & -3.51 & 0.13& -31.00 & 21.67&  2 & 9.50 & 2.60 & 11.90&  2 \\ 
2772 & 306.439 & 0.355 & 0.070 & 0.042 & 15 & 0.27 & 0.03 & -6.83 & 0.09 & -0.78 & 0.07& -74.17 &  &  1 & 8.25 & 3.55 & 12.55&  2 \\ 
2773 & 307.390 & 0.586 & 0.077 & 0.032 & 15 & 0.32 & 0.05 & -6.80 & 0.06 & -1.53 & 0.03&   &  &  0 & 6.00 & 2.35 & 12.70&  2 \\ 
2774 & 78.575 & 5.370 & 0.223 & 0.212 & 15 & 0.31 & 0.04 & -2.86 & 0.13 & -6.07 & 0.09&   &  &  0 & 9.10 & 2.50 & 12.90&  2 \\ 
2775 & 92.050 & -1.141 & 0.012 & 0.012 & 16 & 0.35 & 0.08 & -4.43 & 0.07 & -4.17 & 0.06&   &  &  0 & 6.70 & 4.90 & 12.50&  2 \\ 
... & ... & ... & ... & ... & ... & ... & ... & ... & ... & ... & ... & ... & ... & ...& ...& ... & ...& ...\\ 
2925 & 204.361 & -1.152 & 0.210 & 0.165 & 13 & 0.33 & 0.04 & 0.23 & 0.06 & -0.84 & 0.15& 10.35 &  &  1 & 8.60 & 1.70 & 12.35&  3 \\ 
2926 & 120.889 & 1.658 & 0.100 & 0.079 & 13 & 0.56 & 0.02 & -2.26 & 0.02 & -1.80 & 0.06& -32.81 &  &  1 & 8.65 & 3.30 & 11.20&  3\\

%% file: t2.tex
 1994772031344444160 & 115.163 & -5.979 & 1.01 & 0.42 & 0.09 & -4.88 & 0.09 & -1.63 & 0.09 &   &   &17.64 & 1.35 & 6 & 1271\\ 
 1994772516683575808 & 115.102 & -5.977 & 0.97 & 0.46 & 0.03 & -5.13 & 0.02 & -1.55 & 0.02 &   &   &15.27 & 0.77 & 11 & 1271\\ 
 1994773199575828224 & 115.213 & -5.930 & 1.06 & 0.50 & 0.08 & -4.95 & 0.08 & -1.52 & 0.08 &   &   &17.57 & 1.22 & 10 & 1271\\ 
 1994773822353609472 & 114.999 & -5.962 & 1.02 & 0.60 & 0.09 & -5.20 & 0.08 & -1.63 & 0.08 &   &   &17.56 & 1.19 & 5 & 1271\\ 
 1994776227535261696 & 115.161 & -5.885 & 1.04 & 0.37 & 0.06 & -5.18 & 0.06 & -1.58 & 0.06 &   &   &17.05 & 1.06 & 4 & 1271\\ 
 1994776330614491904 & 115.162 & -5.908 & 0.99 & 0.43 & 0.02 & -5.15 & 0.02 & -1.48 & 0.02 &   &   &14.69 & 0.65 & 11 & 1271\\ 
 1994783133843280384 & 115.325 & -5.998 & 1.04 & 0.30 & 0.07 & -5.05 & 0.06 & -1.66 & 0.06 &   &   &17.09 & 1.08 & 8 & 1271\\ 
 1994781553295335680 & 115.355 & -6.022 & 0.97 & 0.45 & 0.01 & -5.12 & 0.01 & -1.63 & 0.01 & -47.99 & 0.21 &10.03 & 1.33 & 11 & 1271\\ 
 1994782515363544064 & 115.254 & -6.034 & 1.17 & 0.55 & 0.10 & -4.97 & 0.09 & -1.54 & 0.09 &   &   &17.61 & 1.28 & 10 & 1271\\ 
 1994782824605636096 & 115.268 & -6.011 & 0.93 & 0.42 & 0.02 & -5.05 & 0.02 & -1.59 & 0.02 &   &   &14.32 & 0.46 & 11 & 1271\\ 
 1994702147937783296 & 115.024 & -6.349 & 1.03 & 0.51 & 0.10 & -5.12 & 0.09 & -1.69 & 0.10 &   &   &17.99 & 1.64 & 5 & 1271\\ 
... & ... & ... & ... & ... & ... & ... & ... & ... & ... & ... & ... & ... & ...& ... & ...\\ 
 4065823717001007616 & 6.299 & -2.299 & 0.97 & 0.48 & 0.06 & 1.27 & 0.06 & 0.66 & 0.05 &   &   &16.45 & 1.61 & 8 & 1691\\ 
 4065823785720508416 & 6.306 & -2.288 & 1.07 & 0.43 & 0.03 & 1.20 & 0.03 & 0.59 & 0.02 &   &   &14.98 & 1.23 & 8 & 1691\\ 
 4065823987534303232 & 6.316 & -2.274 & 0.61 & 0.42 & 0.02 & 1.41 & 0.02 & 0.60 & 0.01 & 14.69 & 0.44 &11.19 & 2.38 & 2 & 1691\\ 
 4065824060598482944 & 6.322 & -2.258 & 0.93 & 0.49 & 0.08 & 1.37 & 0.08 & 0.72 & 0.07 &   &   &16.93 & 1.78 & 6 & 1691\\ 
 4065824228051149824 & 6.353 & -2.291 & 0.99 & 0.62 & 0.10 & 1.27 & 0.10 & 0.65 & 0.07 &   &   &17.34 & 1.92 & 3 & 1691\\ 
 4065824537288877568 & 6.369 & -2.291 & 0.88 & 0.54 & 0.03 & 1.25 & 0.03 & 0.68 & 0.02 &   &   &14.22 & 1.32 & 8 & 1691\\ 
 4065824644714035712 & 6.388 & -2.277 & 0.94 & 0.46 & 0.04 & 1.33 & 0.04 & 0.61 & 0.03 &   &   &15.32 & 1.24 & 8 & 1691\\ 
 4065871408321742848 & 6.347 & -2.243 & 0.63 & 0.46 & 0.02 & 1.24 & 0.02 & 0.55 & 0.02 & 14.64 & 0.19 &11.01 & 2.49 & 7 & 1691\\ 
 4065871816289108992 & 6.392 & -2.228 & 0.87 & 0.51 & 0.05 & 1.32 & 0.03 & 0.68 & 0.03 &   &   &14.73 & 1.27 & 8 & 1691\\ 
 4065825155765975040 & 6.416 & -2.342 & 0.97 & 0.45 & 0.09 & 1.12 & 0.07 & 0.58 & 0.06 &   &   &16.67 & 1.70 & 5 & 1691\\ 
 4065872095516609536 & 6.333 & -2.191 & 0.92 & 0.43 & 0.05 & 1.39 & 0.05 & 0.64 & 0.04 &   &   &15.98 & 1.42 & 2 & 1691\\ 
 4065826087823028736 & 6.413 & -2.294 & 0.60 & 0.51 & 0.02 & 1.20 & 0.02 & 0.59 & 0.01 & 14.77 & 0.60 &10.98 & 2.12 & 8 & 1691\\ 
 4065826255276033536 & 6.429 & -2.287 & 0.91 & 0.55 & 0.04 & 1.28 & 0.04 & 0.69 & 0.03 &   &   &14.91 & 1.24 & 8 & 1691\\ 
... & ... & ... & ... & ... & ... & ... & ... & ... & ... & ... & ... & ... & ...& ... & ...\\ 
  527017711571049984 & 120.730 & 1.556 & 1.02 & 0.57 & 0.02 & -2.23 & 0.02 & -1.80 & 0.02 &   &   &14.66 & 1.46 & 3 & 2926\\ 
  527021010105990528 & 120.956 & 1.584 & 1.00 & 0.57 & 0.01 & -2.27 & 0.01 & -1.86 & 0.01 & -32.81 & 0.89 &13.60 & 2.33 & 3 & 2926\\ 
  527023965043475456 & 120.801 & 1.605 & 1.00 & 0.56 & 0.03 & -2.23 & 0.03 & -1.84 & 0.03 &   &   &15.77 & 1.58 & 3 & 2926\\ 
  527024652238235264 & 120.867 & 1.635 & 1.00 & 0.54 & 0.02 & -2.27 & 0.02 & -1.74 & 0.02 &   &   &14.97 & 1.44 & 3 & 2926\\ 
  527024755317446912 & 120.861 & 1.646 & 1.06 & 0.56 & 0.05 & -2.29 & 0.05 & -1.71 & 0.05 &   &   &16.80 & 1.99 & 4 & 2926\\ 
  527048841494692864 & 120.889 & 1.719 & 0.97 & 0.54 & 0.06 & -2.29 & 0.07 & -1.68 & 0.07 &   &   &17.35 & 1.96 & 3 & 2926\\ 
  527205384460573568 & 120.714 & 1.566 & 0.95 & 0.56 & 0.08 & -2.24 & 0.08 & -1.84 & 0.09 &   &   &17.78 & 2.05 & 1 & 2926\\ 
  527047295306435200 & 121.030 & 1.782 & 0.96 & 0.54 & 0.03 & -2.21 & 0.02 & -1.83 & 0.03 &   &   &15.53 & 1.69 & 2 & 2926\\ 
  527049563049154688 & 120.875 & 1.774 & 0.99 & 0.56 & 0.09 & -2.25 & 0.09 & -1.84 & 0.09 &   &   &17.94 & 1.93 & 1 & 2926\\ 